\documentclass[12pt]{article}
\usepackage[paperwidth=8.5in,paperheight=11in,hmargin=1in,vmargin={1in,0.5in}
,footskip=0.5in,nohead,dvips]{geometry}

\usepackage{amssymb,amsmath,setspace,theorem}

\newtheorem{lemma}{Lemma}
\newtheorem{theorem}{Theorem}

\newcommand{\qed}{\hfill \ensuremath{\Box}}

\newcommand{\rh}{\mathcal{R}_2}
\newcommand{\R}{\mathcal{R}_1}
\newcommand{\barR}{\bar{\mathcal{R}}_1}
\newcommand{\barrh}{\bar{\mathcal{R}}_2}
\newcommand{\hSp}{\hat{\Sigma}_+}
\newcommand{\hv}{\hat{v}}
\newcommand{\hR}{\hat{R}} 

\newcommand{\W}{\mathcal{W}}
\newcommand{\Wm}{\mathcal{W}_1}
\newcommand{\Wc}{\mathcal{W}_2}

\newcommand{\Sp}{\Sigma_+}
\newcommand{\Sm}{\Sigma_-}
\newcommand{\Sc}{\Sigma_\times}

\newcommand{\Np}{N_+}
\newcommand{\Nm}{N_-}
\newcommand{\Nc}{N_\times}

\newcommand{\Ep}{\mathcal{E}_+}
\newcommand{\Em}{\mathcal{E}_-}
\newcommand{\Ec}{\mathcal{E}_\times}

\newcommand{\Hp}{\mathcal{H}_+}
\newcommand{\Hm}{\mathcal{H}_-}
\newcommand{\Hc}{\mathcal{H}_\times}

\newcommand{\Oml}{\Omega_\Lambda}
\newcommand{\bigO}{\mathcal{O}}

\def\be{\begin{equation}}
\def\ee{\end{equation}}
\def\etal{\emph{et al} }

\begin{document}
\begin{spacing}{1}

\begin{center}
{\large\bf Tilted Bianchi VII$_0$ cosmologies -- the radiation 
bifurcation}
\\[5mm]
W C Lim$^{1,2}$, R J Deeley$^{1,3}$ and J Wainwright$^{1}$
\\[2mm]
$^{1}$ Department of Applied Mathematics, University of Waterloo,\\
Waterloo, Ontario, Canada N2L 3G1\\
$^{2}$ Department of Mathematics and Statistics, Dalhousie University,\\
Halifax, Nova Scotia, Canada B3H 3J5\\
$^{3}$ Department of Mathematics and Statistics, University of Victoria,
Victoria, British Columbia, Canada V8W 3P4\\
[2mm]
Email: wclim@mathstat.dal.ca, rjdeeley@math.uvic.ca, 
	jwainwri@math.uwaterloo.ca\\[2mm]
\end{center}

\begin{abstract}
We derive the late-time behaviour of tilted Bianchi VII$_0$ cosmologies 
with an irrotational radiation fluid as source, and give the asymptotic 
form of the general solution as $t \rightarrow +\infty$, making
comparisons with the dust-filled models.
At first sight 
the radiation-filled
models appear to approximate the flat FL model at late times, since 
the Hubble-normalized shear and the tilt tend to zero and the density 
parameter tends to one. The Hubble-normalized Weyl
curvature diverges, however, 
indicating that physically significant anisotropy remains.
We also discuss the influence of a cosmological constant on this 
phenomenon.
\end{abstract}

\section{Introduction}\label{sec:intro}

The evolution of spatially homogeneous cosmologies of Bianchi type VII$_0$ 
with a non-tilted perfect fluid source 
and zero cosmological constant
has recently been shown to exhibit 
a number of new and interesting features at late times 
(see Wainwright \etal 1999 and Nilsson \etal 2000)%
\footnote{From now on we will refer to these papers as WHU and NHW 
respectively.}
Firstly, they exhibit the \emph{breaking of asymptotic self-similarity}, 
and thus serve as counter-examples to the hope that any Bianchi universe 
is approximated at late times by an exact self-similar Bianchi universe
(see WHU, pages 2578--2579).

The breaking of asymptotic self-similarity is characterized by 
oscillations in the Hubble-normalized shear that become increasingly rapid 
in terms of the cosmological clock time $t$ as $t \rightarrow +\infty$.
Depending on the equation of state, this behaviour also leads to
\emph{Weyl curvature dominance}, which refers to the fact that 
Hubble-normalized scalars formed from the Weyl curvature tensor become 
unbounded as $t \rightarrow +\infty$ (see WHU, page 2588).

Asymptotic self-similarity breaking also has interesting implications as 
regards isotropization. In a spatially homogeneous cosmology, there are 
two physical manifestations of anisotropy:
\begin{itemize}
\item[(a)]	the shear of the timelike congruence that represents the 
		large-scale distribution of the matter;
\item[(b)]	the Weyl curvature, which can be viewed as describing the 
		intrinsic anisotropy in the gravitational field: it 
		determines up to four preferred directions, the principal 
		null directions of the gravitational field.
\end{itemize}
The dynamical significance of these anisotropies can be quantified by 
normalizing the shear tensor and the Weyl curvature tensor with the Hubble 
scalar, and defining two scalars, the 
\emph{shear parameter} $\Sigma$ and the 
\emph{Weyl parameters} $\Wm$, $\Wc$
(see Section~\ref{sec:iso} for the definitions).%
\footnote{We shall use $\Wm$ and $\Wc$ in this paper instead of
$\mathcal{W}$ as in WHU and NWH because $\Wm$ and $\Wc$ are spacetime
scalars while $\mathcal{W}$ is defined with respect to the congruence, and
hence is not a spacetime scalar.}

Bianchi I cosmologies isotropize for all values of the equation of state 
parameter $\gamma-1 = p/\rho$, i.e. $0 < \gamma < 2$, in the sense that 
$\Sigma \rightarrow 0$ and
$\Wm,\Wc \rightarrow 0$ as $t \rightarrow +\infty$.
Indeed these models are asymptotically self-similar, and the asymptotic 
state is the flat Friedmann-Lema\^{\i}tre (FL) model. On the other hand, 
for the class of Bianchi VII$_0$ cosmologies the value $\gamma=\tfrac43$, 
corresponding to a radiation fluid, acts as a 
\emph{bifurcation value}
as regards isotropization:
if $\gamma \leq \tfrac43$, 
the shear parameter $\Sigma \rightarrow 0$ as $t \rightarrow +\infty$,
while
if $\tfrac43 < \gamma < 2$,
$\Sigma$ oscillates indefinitely away from zero, and does not approach a 
limit (see WHU, page 2581).
In addition, the isotropization that occurs in the case
$1 \leq \gamma \leq \tfrac43$
differs significantly from the isotropization that occurs in Bianchi I 
cosmologies, due to the previously mentioned phenomenon of
Weyl curvature dominance.
Specifically, in Bianchi VII$_0$ cosmologies, if $\gamma=1$, the Weyl 
scalars $\Wm$ and $\Wc$ are bounded but do not tend to zero as 
$t \rightarrow +\infty$, 
while if $1 < \gamma \leq \tfrac43$, $\Wm$ and $\Wc$ are unbounded as 
$t \rightarrow +\infty$ (see WHU, page 2581).
We say that shear isotropization occurs if $\gamma \leq \tfrac43$, but 
that Weyl isotropization does not.

More recently, Coley \& Hervik 2005 have investigated the effect of 
tilt on the dynamics of Bianchi VII$_0$ cosmologies at late times.
In general tilted Bianchi cosmologies have three additional degrees of
freedom which describe the orientation of the fluid four-velocity relative
to the normal to the group orbits
(see King \& Ellis 1973).
For simplicity Coley \& Hervik restricted their attention to models with 
one tilt degree of freedom.
They found that the value $\gamma=\tfrac43$ of the equation of state 
parameter acts as a bifurcation value as regards the tilt:
if $\gamma \leq \tfrac43$, the tilt variable tends to zero 
as $t \rightarrow +\infty$
and the model undergoes shear isotropization as in the non-tilted case, 
while if $\tfrac43 < \gamma < 2$, the tilt variable does not tend to zero 
as $t \rightarrow  +\infty$. The analysis of Coley \& Hervik does not 
include models with $\gamma=\tfrac43$, the bifurcation value.
Our first goal in this paper is to give a proof of the asymptotic 
behaviour of this class of cosmologies, i.e. the radiation-filled Bianchi 
VII$_0$ cosmologies with one tilt degree of freedom.
Our second goal is to give the asymptotic decay rates 
as $t \rightarrow +\infty$ for the various physical quantities.

The outline of the paper is as follows.
In Section~\ref{sec:eq} we present the evolution equations for Bianchi 
VII$_0$ cosmologies with a tilted irrotational perfect fluid.
In Section~\ref{sec:limits} we establish the limits of the dimensionless 
gravitational field and matter variables at late times for the case of a 
radiation fluid with zero cosmological constant.
Section~\ref{sec:iso} concerns the question of isotropization at late 
times.
Section~\ref{sec:rates} presents the asymptotic form of the solutions.
Section~\ref{sec:lambda} considers the effect of a positive cosmological 
constant on the results of the previous sections.
We conclude in Section~\ref{sec:discussion} with a discussion of the 
implications of the results.
The details of the proof for the limits are given in 
Appendix~\ref{app:limits}.
The derivation of the asymptotic forms is given in Appendix~\ref{app:N}.

\section{The evolution equations}\label{sec:eq}

As in previous recent investigations of Bianchi VII$_0$ cosmologies
(see WHU, NHW, Coley and Hervik 2005), we write the 
Einstein field equations using Hubble-normalized variables
within the framework of the orthonormal frame formalism of Ellis and 
MacCallum 1969, in which the commutation functions $\gamma^c{}_{ab}$ of 
the orthonormal frame $\{ \mathbf{e}_a \}$, defined by
\[
	[ \mathbf{e}_a,\ \mathbf{e}_b ] = \gamma^c{}_{ab} \mathbf{e}_c \ ,
\]
act as the basic variables of the gravitational field.%
\footnote{Indices are with respect to the basis of the frame vectors 
instead of the coordinate basis; lower case Latin indices run from 0 to 3; 
Greek indices from 1 to 3; upper case Latin indices from 2 to 3.}
We choose the frame to be invariant under the isometry group $G_3$, and
$\mathbf{e}_0 = \mathbf{n}$, the unit normal to the group orbits.
It follows that $\gamma^c{}_{ab}$ depends only on $t$, the clock time 
along $\mathbf{e}_0$.
The non-zero components of the commutation 
functions are (see Wainwright and Ellis, page 39)
\[
	\{ H,\sigma_{\alpha\beta},n_{\alpha\beta},a_\alpha,\Omega_\alpha
	\}\ .
\]
The variable $H$ is the Hubble scalar,
$\sigma_{\alpha\beta}$ is the rate of shear tensor,
$n_{\alpha\beta}$ and $a_\alpha$ describe the spatial curvature, and
$\Omega_\alpha$ is the local angular velocity of the spatial frame
$\{ \mathbf{e}_\alpha \}$ with respect to a Fermi-propagated spatial 
frame.
The energy-momentum tensor in the case of a tilted perfect fluid is
\[
	T_{ab} = \tilde{\rho} u_a u_b + \tilde{p} (g_{ab} + u_a u_b)\ ,
\]
and we assume linear equation of state
$\tilde{p} = (\gamma-1) \tilde{\rho}$.
The fluid 4-velocity $\mathbf{u}$ is written in the form
\[
	u^a = \frac{1}{\sqrt{1-v_b v^b}} (n^a + v^a)\ ,
\]
where the spacelike vector $\mathbf{v}$ is orthogonal to $\mathbf{n}$, and 
satisfies $0 \leq v_b v^b < 1$. The vector $\mathbf{v}$ is called the 
\emph{tilt vector} of the fluid. It has components $(0,v^\alpha)$ relative 
to the orthonormal frame $\{ \mathbf{n}, \mathbf{e}_\alpha \}$.
Writing the energy-momentum tensor in the form
\[
	T_{ab} = \rho n_a n_b + 2 q_{(a} n_{b)} + p (g_{ab} + n_a n_b) + 
		\pi_{ab}\ ,
\]
the source terms $(\rho,p,q_\alpha,\pi_{\alpha\beta})$ 
can be expressed in terms of
$\rho$ and $v_\alpha$ (see Appendix A of Hewitt \etal 2001).
The Hubble-normalized commutation functions and the density paramter 
$\Omega$ are defined by
\be
\label{Hnorm1}
	(\Sigma_{\alpha\beta},N_{\alpha\beta},A_\alpha,R_\alpha)
	= (\sigma_{\alpha\beta},n_{\alpha\beta},a_\alpha,\Omega_\alpha)/H
	\ ,\quad
	\Omega = \rho/(3H^2)\ .
\ee
We introduce the usual dimensionless time variable $\tau$ according to
\be
\label{tau_def}
	\frac{d\, t}{d\tau} = \frac{1}{H}\ .
\ee
For more details, see Appendix A of Hewitt \etal 2001.

For the class of Bianchi VII$_0$ cosmologies with a tilted perfect fluid, 
we have $A_\alpha=0$, and one of the eigenvalues of $N_{\alpha\beta}$ is 
zero. We choose the spatial frame $\{ \mathbf{e}_\alpha \}$ so that
$\mathbf{e}_1$ is the corresponding eigenvector. It then follows that
\be
	N_{1\alpha}=0\ ,
\ee
and the evolution equations of the $N_{1\alpha}$
(Hewitt \etal 2001, equation (A.12))  imply
\be
	R_2 = \Sigma_{13}\ ,\quad
	R_3 = - \Sigma_{12}\ .
\ee
We next impose the restriction that the perfect fluid is
\emph{irrotational}, which is equivalent to%
\footnote{This result follows from the expressions for the fluid vorticity 
$W_\alpha$ in Hewitt \etal 2001, Appendix C. See King \& Ellis 1973 for
the original result.}
setting
\be
\label{irro}
	v_2=v_3=0=\Sigma_{12}=\Sigma_{13}\ .
\ee
We now label the shear variables 
$\Sigma_{\alpha\beta}$ and the spatial curvature variables 
$N_{\alpha\beta}$ as follows
\be
	(\Sigma_{\alpha\beta}) = \left(
	\begin{matrix}
	-2\Sp & 0 & 0
	\\
	0 & \Sp + \sqrt{3} \Sm & \sqrt{3}\Sc
	\\
	0 & \sqrt{3}\Sc & \Sp - \sqrt{3} \Sm
	\end{matrix}
	\right)
	\ ,\quad
        (N_{\alpha\beta}) = \left(
        \begin{matrix}
        0 & 0 & 0
        \\
        0 & \Np + \sqrt{3} \Nm & \sqrt{3}\Nc       
        \\
        0 & \sqrt{3}\Nc & \Np - \sqrt{3} \Nm
        \end{matrix}
        \right)
	\ ,
\label{mat}
\ee
and also drop the index on $v_1$ and $R_1$.

With these restrictions, the general evolution equations (A.11)--(A.13) 
and (A.29) in Hewitt \etal 2001 reduce to the following system of 
differential equations:
\begin{align}
\label{Sp_g}
        \Sp' &= (q-2)\Sp -2(\Nm^2+\Nc^2)
                - G_+^{-1}\gamma \Omega v^2
\\
        \Sm' &= (q-2)\Sm -2R\Sc-2\Np\Nm
\label{Sm_g}
\\
        \Sc' &= (q-2)\Sc +2R\Sm-2\Np\Nc
\label{Sc_g}
\\
        \Np' &= (q+2\Sp)\Np + 6\Sm\Nm + 6\Sc\Nc
\label{Np_g}
\\
        \Nm' &= (q+2\Sp)\Nm - 2R\Nc + 2\Np\Sm
\label{Nm_g}
\\
        \Nc' &= (q+2\Sp)\Nc + 2R\Nm + 2\Np\Sc
\label{Nc_g}
\\
        v' &= G_-^{-1} \bigl[ (3\gamma-4) + 2
                \Sigma_{+} \bigr] (1-v^{2})v \ ,
\label{veq_g}
\end{align}
where
\[   
        G_\pm = 1 \pm (\gamma-1) v^2\ .
\]
The density parameter $\Omega$ is given by
\be
\label{Omega_general}
        \Omega = 1-\Sp^2-\Sm^2-\Sc^2-\Nm^2-\Nc^2 \geq 0\ ,
\ee
and the deceleration parameter $q$ is given by
\begin{equation}
\label{eqs_q}
        q = 2(\Sp^2+\Sm^2+\Sc^2) + \tfrac12 G_+^{-1}
                [(3\gamma-2)+(2-\gamma)v^2]\Omega\ .
\end{equation}
There is one constraint left from (A.27) of Hewitt \etal 2001:
\begin{equation}
\label{constraint_g}
        0 = 2(\Sm\Nc-\Sc\Nm)+ G_+^{-1} \gamma \Omega v\ .
\end{equation}
The evolution equation (A.28) of Hewitt \etal 2001 for $\Omega$ simplifies 
to
\be
        \Omega' = \left[ 2q - G_+^{-1}[(3\gamma-2)+(2-\gamma)v^2]
                	+2G_+^{-1}\gamma \Sp v^2 \right] \Omega\ .
\label{Omega_prime}
\ee
The variable $R$ in equations (\ref{Sp_g})--(\ref{veq_g}), which describes
the rate of which the spatial frame is rotating in the 23-space, is
unrestricted. We will eliminate it by introducing rotation-invariant
variables as in (\ref{inv1})--(\ref{inv3}) below.

The inequality (\ref{Omega_general}) implies that the variables $\Sp$, 
$\Sm$, $\Sc$, $\Nm$ and $\Nc$ are bounded.
The remaining gravitational field variable, $\Np$, however, is not 
bounded.
In fact, it has been shown (see equation (4.2) of Coley and Hervik 
2005) that
\be
\label{Np_lim}
	\lim_{\tau \rightarrow + \infty} \Np = \infty
\ee
for all initial states with $\Omega>0$ and $\tfrac23 < \gamma < 2$.
We thus introduce a new variable
\be
\label{M_def}
        M = \frac{1}{\Np}\ .
\ee     
It can be seen from equations (\ref{Sm_g}) \& (\ref{Nm_g}), and 
(\ref{Sc_g}) \& (\ref{Nc_g})
 that the unboundedness of $\Np$ leads to 
increasingly rapid oscillations in the pairs of variables $(\Sm,\Nm)$ and 
$(\Sc,\Nc)$. 
We thus introduce an angular variable $\psi$ and two rotation-invariant 
variables $\R$ and $\rh$ as follows
\begin{align}
\label{inv1}
	\R &=   \Sigma_-^2 + \Sigma_\times^2 + N_-^2 + N_\times^2
\\
\label{inv2}
	\rh \cos 2\psi &=
		\Sigma_-^2 + \Sigma_\times^2 - N_-^2 - N_\times^2
\\
	\rh \sin 2\psi &= 2(\Sigma_- N_- + \Sigma_\times N_\times)\ .
\label{inv3}
\end{align}
It follows from (\ref{inv1})--(\ref{inv3}) that
\be
\label{invE}
	\R^2 - \rh^2 = 4 (\Sigma_- N_\times - \Sigma_\times N_-)^2\ .
\ee
The resulting evolution equations for the variables 
$(\Sp,\R,\rh,M,\psi,v)$ do not contain the variable $R$.

We note that the conditions
\be
	\Sc=0 \ ,\quad \Nc=0 \ ,\quad R=0\ ,
\label{nt_condition}
\ee
which imply $v=0$ on account of (\ref{constraint_g}), describe the 
non-tilted models. With these restrictions equations 
(\ref{Sp_g})--(\ref{eqs_q}) reduce to (3.7), (3.9) and (3.10) in WHU.
In addition the angular variable $\psi$ in (\ref{inv2}) and (\ref{inv3}) 
coincides with the angular variable $\psi$ defined by equation (3.15) in 
WHU.
This conclusion follows from noting that (\ref{nt_condition}) implies 
$\R=\rh$ and that $\R=R^2_{\rm WHU}$.

\section{Limits for the radiation-filled case at late 
		times}\label{sec:limits}

In this section we determine the limits of the variables
\be
\label{vars}
        (\Sp,\ \R,\ \rh,\ M,\ \psi,\ v)\ .
\ee
for the case of a radiation fluid, given by
\be
	\gamma = \tfrac43\ .
\ee
For this value of $\gamma$, the expression (\ref{eqs_q}) for $q$
simplifies to
\be
\label{q27}
	q = 1 + \Sp^2 + \rh \cos 2\psi\ ,
\ee
on eliminating $\Omega$ using (\ref{Omega_general}).
In terms of the variables (\ref{vars})
the evolution equations (\ref{Sp_g})--(\ref{veq_g}) assume the form%
\footnote{These equations can also be obtained from the evolution 
equations (4.10)--(4.18) in
Coley and Hervik 2005,
by setting $\gamma=\tfrac43$, and 
noting the following relation between our variables and theirs:
\[
        \R = \sigma_1\ ,\quad \rh = \rho\ ,\quad
        2\psi = \tfrac{\pi}{2}- \psi_{\rm CH} \ ,\quad
        M = \tfrac{1}{\sqrt{3}} M_{\rm CH}\ .
\]
}
\begin{align}
\label{Sp}
	\Sigma'_{+} &= -(1-\Sp^2) \Sigma_{+} -\R + (1+\Sigma_{+})
	\rh \cos 2\psi - \frac{4 \Omega v^{2}}{3+v^2}
\\
\label{Req}
	\R' &= 2 \bigl[ (1+\Sp)\Sp + \rh \cos 2\psi
		\bigr] \R - 2 (1+ \Sigma_{+}) (\cos 2\psi) \rh
\\
	\rh'&= 2 \bigl[ (1+\Sp)\Sp + \rh \cos 2\psi
		\bigr]\rh - 2 (1+ \Sigma_{+}) (\cos 2\psi) \R
\\
\label{Meq}
	M' &= - [ (1+\Sp)^2 + \rh \cos 2\psi + 3 \rh M
		\sin 2\psi ] M
\\   
	\psi' &= \tfrac{2}{M} + (1+ \Sigma_{+}) \tfrac{\R}{\rh} \sin 2\psi
\label{psieq}
\\
	v' &= \tfrac{6}{3-v^2} \Sigma_{+} (1-v^{2})v   \ .
\label{veq}
\end{align}
Using (\ref{invE}), the constraint (\ref{constraint_g}) becomes
\be
\label{constraint}
	\R^{2} - \rh^{2} - \left( \frac{4 \Omega v}{3+v^2}
		\right)^{2} = 0\ .
\ee
Equation (\ref{Omega_general}) for the density parameter assumes the form
\be
\label{Friedmann}
	\Omega = 1-\Sigma_{+}^{2} - \R \ .
\ee
The evolution equation (\ref{Omega_prime}) for $\Omega$ becomes
\be
\label{Oeq}
	\Omega' = 2\left[ \Sp^2 + \rh \cos 2\psi + \frac{4v^2}{3+v^2}\Sp 
		\right] \Omega\ .
\ee

Observe that equations (\ref{Sp})--(\ref{Oeq}) are invariant under the
interchange $v \rightarrow - v$, and that $v=0$ is an invariant set. We
can thus without loss of generality assume that $v > 0$ for tilted models.
The above formulation of the evolution equations has one unsatisfactory 
feature, namely the presence of the term $\R/\rh$ in the evolution 
equation (\ref{psieq}) for $\psi$. 
By (\ref{inv1}), $\R \geq 0$ for all $\tau$, 
and on account of (\ref{constraint}), 
$\R=0$ for some $\tau$ is not possible 
since then $\Omega v =0$, which is excluded for tilted matter-filled 
models.
So $\R>0$ for all $\tau$, i.e. $\R=0$ is an invariant set, in the boundary 
of the physical state space.
On the other hand, $\rh$ can become zero, in which case the change of 
variables (\ref{inv2})--(\ref{inv3}) is singular and the evolution 
equations (\ref{Sp})--(\ref{veq}) break down.
Numerical simulations using the evolution equations 
(\ref{Sp_g})--(\ref{veq_g}), however, provide evidence that for typical 
initial conditions $\rh$ can become zero for at most a finite number of 
times.
In other words for a typical choice of initial condition, there exists a 
time $\tau_0$ such that $\rh(\tau)>0$ or $\rh(\tau)<0$ for all $\tau 
> \tau_0$.
Note that the evolution equations
(\ref{Sp})--(\ref{veq})
are invariant under the transformation 
$\psi \rightarrow \psi + \frac{\pi}{2}$ and $\rh \rightarrow -\rh$. We
make use of this symmetry to assume, without loss of generality, that 
$\rh(\tau)>0$  for all $\tau > \tau_0$.
%In addition the simulations provide evidence that for a typical initial 
%condition, the ratio $\R/\rh$ is bounded as $\tau \rightarrow +\infty$.
In other words, for typical initial conditions we expect that for $\tau$ 
sufficiently large, the transformation (\ref{inv1})--(\ref{inv3}) will be 
non-singular, and the evolution equations (\ref{Sp})--(\ref{veq}) will be 
well-defined.

The limit of $M$ is
\be
\label{M_lim}
	\lim_{\tau \rightarrow +\infty} M =0\ ,
\ee
which follows immediately from (\ref{Np_lim}).
Since $\psi'$ in (\ref{psieq}) is dominated by $M^{-1}$,
the trigonometric functions in (\ref{Sp})--(\ref{veq}) will oscillate 
increasingly rapidly as $\tau \rightarrow +\infty$.
In order to control these oscillations, we
modify the variables $\Sp$, $\R$, $M$, $Z$ and $\Omega$
(in analogy with equations (17)--(21) in NHW) according to
\begin{align}
\label{Spbar}
        \bar{\Sigma}_{+} & = \Sigma_{+}
                -\tfrac{1}{4}M(1+\Sigma_{+}) \rh \sin 2\psi
\\
\label{Rbar}
        \barR & = \frac{\R}{1
                +\frac{1}{2}M(\R-1-\Sigma_{+})\frac{\rh}{\R}\sin 2\psi}
\\
\label{Mbar}
        \bar{M} & = \frac{M}{1-\frac{1}{4} M \rh \sin 2\psi}
\\
\label{Ombar}
        \bar{\Omega} & = \frac{\Omega}{1 + \frac{1}{2} M \rh \sin 2\psi}
				\ .
\end{align}
It follows from (\ref{Sp}), (\ref{Req}), (\ref{Meq}), (\ref{Friedmann}) 
and (\ref{Oeq})
that the evolution equations for the new variables have the form%
\footnote{We have chosen to write equation (\ref{bar_Omega}) with $\Sp$
unbarred in the second term, since this form is required in the proof
in Appendix~\ref{app:limits}.
The factor of $\R$ in front 
of $B_{\bar{\Omega}}$ is crucial, and $\rh$ has to be written as 
$(\frac{\rh}{\R})\R$ to produce this factor. We recall from 
(\ref{constraint}) that $-1 \leq \frac{\rh}{\R} \leq 1$.}
\begin{align}
\label{bar_Sp}
        \bar{\Sigma}_{+}^{\prime} & =
        -\barR-\bar{\Sigma}_{+}(1-\bar{\Sigma}_{+}^{2})
	-\frac{4}{3+v^2}(1-\bar{\Sigma}_+^2 -\barR)v^2
        + M B_{\bar{\Sigma}_{+}}
\\
\label{bar_R}
        \barR^{\prime} & = 2\left[\bar{\Sigma}_{+}
        +\bar{\Sigma}_{+}^{2}+M B_{\barR}\right]\barR
\\
\label{bar_M}
        \bar{M}^{\prime} & = -\left[(1+\bar{\Sigma}_{+})^2
        +M B_{\bar{M}}\right]\bar{M}
\\
\label{bar_Omega}
        \bar{\Omega}^{\prime} & = 2\left[\bar{\Sigma}_{+}^2
        +\frac{4 \Sp v^2}{3+v^2}+M \R B_{\bar{\Omega}}\right]
        \bar{\Omega}\ ,
\end{align}
where the $B$'s are bounded as $\tau \rightarrow +\infty$.
Equations (\ref{bar_Sp})--(\ref{bar_Omega}) depend explicitly on $\psi$ 
through the $B$'s which are all multiplied by a factor $M$. This fact, in 
conjunction with (\ref{M_lim}), implies that 
\emph{the oscillatory terms in (\ref{bar_Sp})--(\ref{bar_Omega}) tend to 
zero as $\tau \rightarrow +\infty$}.
Observe that the variable $\rh$ only appears in the $B$-terms, which means
that it is not necessary to introduce an evolution equation for a modified
variable $\barrh$.
The evolution equation (\ref{veq}) for $v$ remains unchanged. 

The first step in the analysis is to use the evolution equations 
(\ref{Sp}) and (\ref{veq}) for $\Sp$ and $v$ to prove that 
$\lim_{\tau \rightarrow +\infty} v = 0$.
The structure of (\ref{bar_Sp})--(\ref{bar_Omega}) then allows the limits 
of $\bar{\Sigma}_+$ and $\barR$ as $\tau \rightarrow +\infty$ to be 
determined, and subsequently the corresponding limits of $\Sp$, $\R$ and
$\rh$.
The main result is contained in the following theorem.

\begin{theorem}\label{thm1}
Any solution of the system (\ref{Sp})--(\ref{constraint}) with $\Omega>0$, 
$\R>0$ and $\rh>0$ satisfies
\be
\label{thm1_1}
	\lim_{\tau \rightarrow +\infty} 
	(\Sigma_+,\R,\rh,v,M) = (0,0,0,0,0),
\ee
with
\be
\label{thm1_2}
        \lim_{\tau \rightarrow +\infty} \frac{\R}{M^2} = + \infty
	\ ,\quad
        \lim_{\tau \rightarrow +\infty} \frac{\rh}{M^2} = + \infty  \ .
\ee
\end{theorem}

\paragraph{Proof.}
The details of the proof are given in Appendix~\ref{app:limits}.   
\qed

\

\section{Isotropization of radiation-filled models at late 
	times}\label{sec:iso}

In this section we will consider the physical implications of 
Theorem~\ref{thm1} as regards isotropization of the radiation-filled
cosmological models with zero cosmological constant.
As mentioned in the introduction, the anisotropy can be quantified by the 
shear parameter $\Sigma$ and the Weyl parameters.
The shear parameter is defined by $\Sigma^2 =
\frac{\sigma_{\alpha\beta}\sigma^{\alpha\beta}}{6H^2}$, or 
equivalently, on account of (\ref{Hnorm1}), by
\be
\label{shear_scalar_def}
	\Sigma^2 = \tfrac16 \Sigma_{\alpha\beta}\Sigma^{\alpha\beta}
	\ .
\ee
It follows from (\ref{mat}) and (\ref{shear_scalar_def}) that
\be
\label{shear_scalar}
	\Sigma^2 = \Sp^2 + \Sigma_-^2 + \Sigma_\times^2\ .
\ee
We define the Weyl parameters $\Wm$ and $\Wc$ by
\be
\label{WmWc_1}
	\Wm = \frac{C_{abcd} C^{abcd}}{48H^4}
	\ ,\quad
	\Wc = \frac{C_{abcd}{}^* C^{abcd}}{48H^4}\ ,
\ee
or equivalently,
\be
\label{WmWc_2}
	\Wm = \frac{E_{\alpha\beta} E^{\alpha\beta} - 
		H_{\alpha\beta} H^{\alpha\beta}}{6H^4}
	\ ,\quad
	\Wc = \frac{E_{\alpha\beta} H^{\alpha\beta}}{3H^4}
        \ ,
\ee
where $E_{\alpha\beta}$ and $H_{\alpha\beta}$ are respectively the 
electric and magnetic parts of the Weyl curvature tensor with respect to 
the congruence orthogonal to the $G_3$ group orbits
(Wainwright and Ellis 1997, page 19).
We define the Hubble-normalized components by
\be
\label{Hnorm_Weyl}
	\mathcal{E}_{\alpha\beta} = \frac{E_{\alpha\beta}}{H^2}
	\ ,\quad
        \mathcal{H}_{\alpha\beta} = \frac{H_{\alpha\beta}}{H^2}
	\ ,
\ee
labelling them
as with $\Sigma_{\alpha\beta}$ in (\ref{mat}).
It follows that
\begin{align}
\label{Weyl1}
	\Wm &= \Ep^2 + \Em^2 + \Ec^2 - \Hp^2 - \Hm^2 - \Hc^2
\\
	\Wc &= 2(\Ep \Hp + \Em \Hm + \Ec \Hc)
	\ .
\label{Weyl3}
\end{align}
They are given by%
\footnote{See Wainwright and Ellis 1997, page 35 for the general
expression for $E_{\alpha\beta}$ and $H_{\alpha\beta}$. See Lim 2004, page
49 for the expression for the individual components, but note that
the components are normalized by $3H^2$ there.}
\begin{align}
\label{Ep}
        \Ep &=  (1+\Sp)\Sp - (\Sm^2+\Sc^2)
                + 2(\Nm^2+\Nc^2)
                + \tfrac12 G_+^{-1}\gamma\Omega v^2
\\
        \Em &= (1-2\Sp)\Sm + 2 \Np \Nm
\\
        \Ec &= (1-2\Sp)\Sc + 2 \Np \Nc
\\
        \Hp &= - 3 (\Sm \Nm + \Sc \Nc)
\\ 
        \Hm &= - 3 \Sp \Nm - 2 \Np \Sm
\\
        \Hc &= - 3 \Sp \Nc - 2 \Np \Sc\ .
\label{Hc}
\end{align}
The limits of the anisotropy scalars $\Sigma$, $\Wm$ and $\Wc$ as
$\tau \rightarrow +\infty$ for tilted irrotational radiation-filled 
Bianchi VII$_0$ cosmologies are given in the following theorem.

\begin{theorem}\label{thm2}
For any tilted irrotational radiation-filled Bianchi VII$_0$ cosmology
with zero cosmological constant,
satisfying $\Omega>0$, $\R>0$, $\rh>0$ and $v>0$, the density parameter 
satisfies
\be
\label{thm2_eq1}
        \lim_{\tau \rightarrow +\infty} \Omega = 1\ ,
\ee
the tilt variable $v$ satisfies
\be
\label{thm2_eq2}
	\lim_{\tau \rightarrow +\infty} v = 0\ ,
\ee
and the anisotropy scalars satisfy
\be
\label{thm2_eq3}
	\lim_{\tau \rightarrow +\infty} \Sigma = 0\ ,\quad
	(\Wm,\Wc) = -\frac{4\rh}{M^2} \left[ (\cos 2\psi,\sin 2\psi)
		+ \bigO(M) \right] \ ,
\ee
with
\be
\label{thm2_eq4}
	\lim_{\tau \rightarrow +\infty}  \frac{\rh}{M^2} = +\infty\ .
\ee
\end{theorem}

\paragraph{Proof.}
It follows from (\ref{inv1}), (\ref{inv2}) and (\ref{shear_scalar}) that
\be
\label{Sigma_thm2}
	\Sigma^2 = \Sp^2 + \tfrac12(\R + \rh \cos 2 \psi)\ .
\ee
Equations (\ref{thm2_eq1}), (\ref{thm2_eq2}) and (\ref{thm2_eq3}) now
follow directly from Theorem~\ref{thm1} and equation 
(\ref{Friedmann}) for $\Omega$.
The asymptotic forms for $\Wm$ and $\Wc$ follow directly from
(\ref{Weyl1})--(\ref{Hc}) and the definitions (\ref{M_def})--(\ref{inv3})
of $M$, $\R$ and $\rh$, on noting that the terms containing $\Np$ are the
dominant ones.
The limit (\ref{thm2_eq4}) is part of Theorem~\ref{thm1}.
\qed

\

The limits in Theorem~\ref{thm2} show that 
\emph{tilted irrotational radiation-filled Bianchi VII$_0$ cosmologies 
isotropize with respect to the shear, but not with respect to the Weyl 
curvature}.

\section{Asymptotic solution at late times}\label{sec:rates}

In this section we derive the asymptotic form 
as $\tau \rightarrow +\infty$ of the basic variables
$(\Sp,\ \R,\ \rh,\ M,\ v)$, which determine the asymptotic form of the 
anisotropy scalars.
We begin by considering the so-called truncated differential equations for 
the variables $\Sp$, $\R$ and $v$, obtained from (\ref{bar_Sp}), 
(\ref{bar_R}) and (\ref{veq}) by setting to zero the terms that involve 
$M$. We also write $\R=R^2$, since this facilitates the analysis.
The resulting differential equations are
\begin{align}
	\hat{\Sigma}_+' &= - \hat{\Sigma}_+ + \hat{\Sigma}_+^3 - \hat{R}^2
	- \frac{4}{3+\hat{v}^2}(1-\hat{\Sigma}_+^2-\hat{R}^2)\hat{v}^2
\label{dSphat}
\\
	\hat{R}' &= \hat{\Sigma}_+ (1+\hat{\Sigma}_+)\hat{R}
\label{dRhat}
\\
	\hat{v}' &= \frac{6}{3-\hat{v}^2}(1-\hat{v}^2) \hat{\Sigma}_+ 
			\hat{v}\ ,
\label{dvhat}
\end{align}
where we use a hat to denote the solutions of the truncated differential 
equations.
In Appendix~\ref{app:N} we show that the solutions of 
(\ref{dSphat})--(\ref{dvhat}) exhibit a power law decay to zero.
Since $M$ decays to zero exponentially fast (see (\ref{M151})) it is
plausible that the 
solutions $\bar{\Sigma}_+$, $\barR$ and $v$ of the full system 
(\ref{bar_Sp}), (\ref{bar_R}) and (\ref{veq}) will have the same 
asymptotic form as the solutions of the truncated differential equations
(\ref{dSphat})--(\ref{dvhat}).
Finally (\ref{Spbar}) and (\ref{Rbar}) show that $\Sp$ and $\R$ have the 
same decay rates as $\bar{\Sigma}_+$ and $\barR$.
An asymptotic expansion for $M$ can most easily be obtained by using the
auxiliary variable $Z$, defined by (\ref{Zdef}), with $\Omega$ given by
(\ref{Friedmann}).
Knowing the asymptotic expansions for $v$ and $\Sp$, and that $M$ decays
exponentially (see (\ref{M151})), we can integrate (\ref{zp}) to obtain an
asymptotic expansion for $\bar{Z}$ and hence recover $Z$ from
(\ref{barZ_def}). 
This process introduces another arbitrary constant $C_M$.

These considerations lead to the following asymptotic form of the basic
variables in any tilted Bianchi VII$_0$ cosmology with
irrotational radiation fluid.
We give the asymptotic expansion for $\R$ explicitly in terms of $\tau$,
and for the remaining variables in terms of $\R$: 
\begin{align}
\label{R1_expansion}
	\R &= \frac{1}{2\tau}
	\left[ 1 - \frac{1}{2\tau} (1+\tfrac43 C_v^2)(\ln \tau + C_{\R})
		+ \bigO \bigg( \Big( \frac{\ln \tau}{\tau} \Big)^2 \bigg) 
	\right]_\frac{}{} ,
\\
\label{v_expansion}
	v &= C_v \R \left[ 1 + \R + \bigO(\R^2) \right]^{} \ ,
\frac{}{}\\
\label{Sp_expansion}
	\Sp &= - \R \left[ 1 + 2(1+\tfrac23 C_v^2)\R + \bigO(\R^2) \right]
		\ ,
\frac{}{}\\
\label{M_expansion}
	M^{-1} &= C_M e^\tau \R \left[ 1 + \tfrac12 \R + \bigO(\R^2)
		\right] \ ,\frac{}{}
\end{align}
as $\tau \rightarrow +\infty$.
Here $C_M$, $C_v$ and $C_{\R}$ are
constants that depend on the initial conditions.
The asymptotic expansion for $\rh$ can be obtained directly from
(\ref{constraint}) and (\ref{Friedmann}). Its leading term is
\be
\label{R2_expansion}
	\rh = \sqrt{1-\tfrac{16}{9} C_v^2}\ \R [ 1+ \bigO(\R) 
                ] \ ,
\ee
which entails the restriction
$0 < C_v \leq \frac{3}{4}$.

We note that the asymptotic expansion for $\Sp$
also contains a term $C_{\Sp} e^{-\tau}$,
and the asymptotic expansion for $\psi$ would contain an
additive constant $C_\psi$. 
Thus all five constants of integration for the
five-dimensional system (\ref{Sp})--(\ref{constraint}) have been accounted
for, confirming that the asymptotic form above describes the general
Bianchi VII$_0$ solution with a radiation fluid source. 
We can set one of these constants to unity by using the freedom to shift
$\tau \rightarrow \tau + C$, leaving four essential constants to
parametrize the four-dimensional family of orbits in the five-dimensional
state space.
We also note the flawed argument in NHW (see page 3126), which should be
corrected as above.

We now give the asymptotic dependence of the physical quantities of
primary interest on the cosmological clock time $t$, which is related to
$\tau$ via equation (\ref{tau_def}).
%The Hubble variable $H$ can be determined
%algebraically through the relation $\Omega = \rho / (3H^2)$, since,
%for a radiation fluid, the quantity $\rho \ell^4$ is a constant, where
%$\ell = C e^\tau$ is the length scale function%
%\footnote{The latter relation follows from (\ref{tau_def}) since
%$H = \frac{1}{\ell}\frac{d\ell}{dt}$ (Wainwright \& Ellis 1997, pages
%112--3).}.
In the present situation the Raychaudhuri equation reads
\be
	H' = -(1+q)H\ ,
\ee
(see Hewitt \etal 2001, equation (A.9)). On account of (\ref{q27}), it
assumes the form
\be
	H' = -(2+\Sp^2 + \rh \cos 2\psi)H \ .
\ee
We integrate this equation using (\ref{R1_expansion}),
(\ref{Sp_expansion}) and (\ref{R2_expansion}),
which gives%
\footnote{Note that the integration of the exponentially rapidly
oscillating term $\rh \cos 2\psi$ yields a constant.}
\be
        H = \tfrac12 A e^{-2\tau} \left[ 1 +
				\bigO(\tfrac{1}{\tau})
        \right] \ ,
\ee
where $A$ is a constant with dimensions of (time)$^{-1}$,
and then use (\ref{tau_def}) to obtain
\be
	t = \frac{1}{A} e^{2\tau} \left[ 1 +
				\bigO(\tfrac{1}{\tau}) 
	\right] \ .
\ee
Note that on account of (\ref{Friedmann}), (\ref{R1_expansion}) and
(\ref{Sp_expansion}),
\be
	\Omega = 1 + \bigO(\tfrac{1}{\tau})\ ,
\ee
as $\tau \rightarrow + \infty$. The energy density $\rho$ can then be
determined
algebraically through the relation $\Omega = \rho / (3H^2)$.

The desired asymptotic forms, valid for $At \gg 1$, are as follows:
\begin{align}
\label{Hrhov_rad}
	H &\approx \frac1{2t}
	\ ,\quad
	\rho \approx \frac{3}{4t^2}
	\ ,\quad
	v \approx \frac{C_v}{\ln(At)}\ ,
\\
\label{Sig_rad}
	\Sigma^2 &\approx \frac{1}{2\ln(At)} 
		\left[1 + \sqrt{1-\tfrac{16}{9}C_v^2}\, 
		\cos 2\psi\right]_\frac{}{} ,
\\
\label{WmWc_rad}
	(\Wm,\Wc) &\approx
	 -\frac{4 C_M^2 (At) }{[\ln(At)]^3} 
		\sqrt{1-\tfrac{16}{9}C_v^2}\ (\cos 2\psi,\sin 2\psi)
		\ ,
\intertext{where}
\label{psi_rad}
	\frac{d\psi}{dt} &\approx \frac{C_M A}{\sqrt{At}\ln(At)}\ .
\end{align}

Observe that the tilt constant $C_v$ primarily affects the amplitude of
the shear oscillations through (\ref{Sig_rad}), while the constant $C_M$
affects how rapidly the shear oscillates through (\ref{psi_rad}), and
hence the amplitude of the Weyl curvature scalars $\Wm$ and $\Wc$.

Although our emphasis in this paper is on the radiation-filled models 
($\gamma=\tfrac43$), we also give the asymptotic forms
as $\tau \rightarrow +\infty$ for the models with
$\tfrac23 < \gamma < \tfrac43$, which includes the physically important 
case 
of dust ($\gamma=1$), in order to make comparisons.
The derivation of these results is an extension of the analysis given in
Appendix B of WHU.

For any tilted irrotational perfect fluid
Bianchi VII$_0$ cosmology with $1 \leq \gamma < \tfrac43$,
\begin{align}
	v &= C_v e^{-2\beta\tau} \left[ 1 + \bigO(e^{-b\tau}) \right]\ ,
\\
        \R &= C_{\R} e^{-2\beta\tau} \left[ 1 + \bigO(e^{-b\tau}) 
	\right]\ ,
\\
	\rh &= \sqrt{ C_{\R}^2 - \gamma^2 C_v^2 } \, 
		e^{-2\beta\tau}[1+\bigO(e^{-b\tau})]\ ,
\\
	\Sp &= - \frac{ C_{\R}}{1-\beta} e^{-2\beta\tau}
			\left[ 1 + \bigO(e^{-b\tau}) \right]\ ,
\\
	M^{-1} &= C_M e^{(1-\beta)\tau} \left[ 1 + \bigO(e^{-b\tau})
		\right]\ ,
\end{align}
as $\tau \rightarrow +\infty$, where
$\beta = \tfrac12(4-3\gamma)$, and
 $C_M$, $C_v$ and $C_{\R}$ are 
constants that depend on the initial conditions,
with $0 < C_v \leq C_{\R}/\gamma$,
 and $b$ is a positive constant.
We note that the asymptotic expansion of $\Sp$
also contains a term $C_{\Sp} e^{-\frac32(2-\gamma)\tau}$,
and the asymptotic expansion for $\psi$ would contain an
additive constant $C_\psi$, giving five constants of integration as in the
radiation case.

We now express the asymptotic behaviour of the various physical quantities
in terms of clock time $t$.
As in the radiation case we obtain
\be
	H = \frac{2A}{3\gamma} e^{-\frac32\gamma\tau}
	\left[ 1 + \bigO(e^{-2\beta\tau}) \right]
	\ ,\quad
	t = \frac{1}{A} e^{\frac32 \gamma \tau} \left[ 1 +
	\bigO(e^{-2\beta\tau}) \right]\ ,
\ee
where $A$ is a constant with dimensions of (time)$^{-1}$.

The desired asymptotic forms, valid for $At \gg 1$, are as follows:
\begin{align}
	H &\approx \frac{2}{3\gamma t}
	\ ,\quad
	\rho \approx \frac{4}{3\gamma^2 t^2}
	\ ,\quad
        v \approx C_v (At)^{-\frac{2(4-3\gamma)}{3\gamma}}\ ,
\\
\label{Sv_subrad}
	\Sigma^2 &\approx \tfrac{1}{2}(At)^{-\frac{2(4-3\gamma)}{3\gamma}}
	\left(C_{\R} + \sqrt{C_{\R}^2 - \gamma^2 C_v^2}\, \cos 2\psi
	\right)\ ,
\\
\label{WmWc_subrad}
	(\Wm,\Wc)
 	&\approx - C_M^2 \sqrt{C_{\R}^2-\gamma^2 C_v^2}\, 
	(At)^\frac{4(\gamma-1)}{\gamma} (\cos 2\psi,\sin 2\psi)\ ,
\intertext{where}
	\psi &\approx \frac{2 C_M}{3\gamma-2} 
	(At)^{\frac{3\gamma-2}{3\gamma}}.
\end{align}

\section{The role of the cosmological constant}\label{sec:lambda}

In this section we discuss the influence of a cosmological constant 
$\Lambda > 0$ on the dynamics of tilted irrotational perfect fluid Bianchi 
VII$_0$ cosmologies.
The evolution equations in Section~\ref{sec:eq} are modified only by the 
presence of 
\be
	\Oml \equiv \frac{\Lambda}{3H^2}
\ee
in (\ref{Omega_general}) and 
(\ref{eqs_q}) as follows:
\begin{align}
        \Omega &= 1-\Sp^2 - \Sm^2-\Sc^2-\Nm^2-\Nc^2 - \Oml
\\
        q &= 2(\Sp^2+\Sm^2+\Sc^2) + \tfrac12 G_+^{-1}
                [(3\gamma-2)+(2-\gamma)v^2]\Omega-\Oml\ ,
\end{align}
and $\Oml$ satisfies the evolution equation
\be
        \Oml' = 2(q+1)\Oml\ .
\label{Oml_g}
\ee
The expression for the Weyl curvature components is not affected.

Wald's theorem (see Wald 1983) makes a strong and general statement about 
the 
asymptotic behaviour of spatially homogeneous cosmologies with $\Lambda>0$
 and subject to the stress-energy tensor satisfying a physically 
reasonably energy condition.
The result is that the models are asymptotic to the de Sitter solution, in 
the sense that 
\be
\label{Oml_deSitter}
	\lim_{\tau \rightarrow +\infty} \Oml = 1 \ ,
\ee
in our notation.
It follows immediately from (\ref{Oml_deSitter}), (\ref{Omega_general}), 
(\ref{shear_scalar}) and (\ref{eqs_q}) that
\be
\label{deSitter_lims}
	\lim_{\tau \rightarrow +\infty} 
	(\Sp,\ \Sm,\ \Sc,\ \Nm,\ \Nc) = 0\ ,
\ee
and hence that
\be
\label{Osq_deSitter}
	\lim_{\tau \rightarrow +\infty} \Omega = 0 \ ,\quad
	\lim_{\tau \rightarrow +\infty} \Sigma = 0 \ ,\quad
	\lim_{\tau \rightarrow +\infty} q = -1 \ .
\ee
Wald's theorem gives no information about the Weyl parameters 
or about the tilt $v$.
However it follows from (\ref{Osq_deSitter}) and (\ref{Np_g}) that
\be
	\lim_{\tau \rightarrow +\infty} \Np = 0 \ ,
\ee
which implies, on account of (\ref{Weyl1})--(\ref{Hc}), that
\be
	\lim_{\tau \rightarrow +\infty} \Wm = 0 \ ,\quad
	\lim_{\tau \rightarrow +\infty} \Wc = 0 \ .
\ee
A significant difference occurs as regards the tilt.
It follows from Section III of Lim \etal 2004 that
\be
	\lim_{\tau \rightarrow +\infty} v
	= \begin{cases}
		0 & \text{if $\tfrac23 < \gamma < \tfrac43$}
	\\
		v_\infty & \text{if $\gamma = \tfrac43$} \ ,
	\end{cases}
\ee
where $0 \leq v_\infty \leq 1$, i.e. the value of $\gamma$ for a radiation
fluid ($\gamma=\frac43$) is a bifurcation value for the tilt in a model
with a cosmological constant.

\section{Discussion}\label{sec:discussion}

In this paper we have derived the late-time dynamical behaviour of tilted
Bianchi VII$_0$ cosmologies with an irrotational radiation fluid as
source, and have given the asymptotic form of the general solution as $t
\rightarrow +\infty$.

The main result is Theorem~\ref{thm1}, which establishes rigorously the
asymptotic behaviour of the cosmological models.%
\footnote{The assumption $\rh>0$ of Theorem~\ref{thm1} is supported 
numerically (see paragraph before (\ref{M_lim})).} 
Theorem~\ref{thm2}, which is an immediate consequence of
Theorem~\ref{thm1}, describes the degree to which the models isotropize.
These theorems generalize the results of NHW for the
corresponding non-tilted models.
The derivation of the detailed asymptotic form of the solutions as
$t \rightarrow +\infty$, given in Section~\ref{sec:rates},
is not completely rigorous, since it relies on the
analysis of a so-called truncated system of differential equations, as
in NHW.

This class of models displays several noteworthy features. 
Firstly, 
the models appear to approximate the flat FL model at late times, since
the Hubble-normalized shear and the tilt tend to zero, and the density
parameter tends to one.
The Hubble-normalized Weyl curvature scalars $\Wm$ and $\Wc$ diverge,
however, indicating that physically significant anisotropy remains.
In effect, the models isotropize as regards shear and tilt, but not as
regards Weyl curvature.
The scalars $\Wm$, $\Wc$, which are defined by equations (\ref{WmWc_1})
and (\ref{WmWc_2}), diverge and oscillate increasingly rapidly in an
out-of-phase manner (see equation (\ref{WmWc_rad})).
The scalar $\W$, defined by
\be
\label{W_def}
	\W^2 = \Wm^2 + \Wc^2,
\ee
then satisfies
\be
\label{W_inf}
	\lim_{t \rightarrow +\infty} \W = + \infty\ ,
\ee
as follows from (\ref{thm2_eq3}) and (\ref{thm2_eq4}).
Thus, in terms of Hubble-normalized scalars, the future asymptotic state
($t \rightarrow +\infty$) for a radiation-filled universe is given by
\be
\label{rad_state}
	\Sigma \rightarrow 0 \ ,\quad
	v \rightarrow 0 \ ,\quad
	\Omega \rightarrow 1 \ ,\quad
	\W \rightarrow \infty\ .
\ee
Secondly,
the isotropization as regards shear and tilt occurs at a slow rate,
i.e. $\Sigma^2$ and $v$ tend to zero logarithmically in terms of clock
time $t$
(proportional to $\frac{1}{\ln(At)}$, see equations (\ref{Hrhov_rad}) and 
(\ref{Sig_rad})).

The dust-filled ($\gamma=1$) universes differ significantly.
The Weyl curvature scalar $\W$ remains bounded and approaches a finite
value $\W_\infty$ that depends on the initial conditions (see equation
(\ref{WmWc_subrad}) with $\gamma=1$).
Thus, in terms of Hubble-normalized scalars, the future asymptotic state
for a dust-filled universe is given by
\be
\label{dust_state}
        \Sigma \rightarrow 0 \ ,\quad
        v \rightarrow 0 \ ,\quad
        \Omega \rightarrow 1 \ ,\quad
        \W \rightarrow \W_\infty\ .
\ee
The isotropization as regards shear and tilt occurs at a faster rate, i.e.
 $\Sigma^2$ and $v$ tend to zero like $(At)^{-2/3}$
(see equation (\ref{Sv_subrad}) with $\gamma=1$).
We note in passing that a universe described by (\ref{dust_state}), i.e.
one that is highly isotropic as regards the shear but not as regards the
Weyl curvature, is compatible with the observed highly isotropic
temperature of the cosmic microwave background radiation (see WHU, Section
4, for further discussion of this point).%
\footnote{Dominance of the Weyl curvature may lead to other observable
effects, such as image distortion (see, e.g., Chrobok \& Perlick 2001.}

As shown in Section~\ref{sec:rates}, the dominance of the Weyl curvature
at late times is curbed by the presence of a positive cosmological
constant (see equation (\ref{deSitter_lims})), i.e. the Weyl scalar $\W$
tends to zero.
Thus, in terms of Hubble-normalized scalars, the future asymptotic state
($t \rightarrow + \infty$) for a universe with a cosmological constant
and radiation or dust, is
\be
        \Sigma \rightarrow 0 \ ,\quad
        v \rightarrow v_\infty \ ,\quad
	\Oml \rightarrow 1 \ ,\quad
        \Omega \rightarrow 0 \ ,\quad
        \W \rightarrow 0\ ,
\ee
with $v_\infty=0$ for dust and $0 \leq v_\infty \leq 1$ for radiation.
The asymptotic states (\ref{rad_state}) or (\ref{dust_state}) with
$\Lambda=0$ are still of relevance for models with $\Lambda>0$ since they
can act as intermediate states, 
i.e. \emph{a model with $\Lambda>0$ can be
approximated arbitrarily closely by the states (\ref{rad_state}) or
(\ref{dust_state}) over a finite time interval}.

In conclusion, we note that the most striking feature of the late-time
dynamics of Bianchi VII$_0$ cosmologies with a radiation fluid is the
so-called \emph{Weyl curvature dominance}: if the cosmological constant is
zero ($\Lambda=0$) then the Hubble-normalized Weyl scalar $\W$ diverges as
$t \rightarrow +\infty$, while if $\Lambda>0$, the scalar $\W$ can become
arbitrarily large during a finite time interval.
The phenomenon was first encountered for non-tilted Bianchi VII$_0$
models (see WHU), 
and in this paper we have shown that it persists in the presence
of tilt, for an irrotational fluid.
The same phenomenon has recently been found in Bianchi VII$_0$ models
with a tilted and rotating fluid (Hervik \etal 2006).
Weyl curvature dominance also occurs in non-tilted Bianchi VIII
cosmologies (Horwood \etal 2003), and has recently been found in tilted
Bianchi VIII cosmologies (Hervik \& Lim 2006).

\section*{Acknowledgements}

We thank Sigbj{\o}rn Hervik, Robert van den Hoogen, Alan Coley and Henk
van Elst for discussions. This work was supported by NSERC through an
Undergraduate Student Research Award (RJD) and a Discovery Grant (JW).

\appendix

\section{Proof of Theorem~\ref{thm1}}\label{app:limits}

In the proof we need the following standard result and a second lemma.

\begin{lemma}\label{lem_Lip}
Let $f(\tau)$ be a non-negative real-valued and Lipschitz continuous
function on the interval $\tau_0 \leq \tau < + \infty$. If
\[
        \int_{\tau_0}^{\infty} f(\tau) \ d\tau \quad\text{is finite}
\]
then $\lim_{\tau \rightarrow +\infty} f(\tau)=0$.
\end{lemma}

\begin{lemma}\label{lem1}
For solutions of equations (\ref{Sp})--(\ref{Oeq}),
the following limits cannot both hold:
\be
        \lim_{\tau \rightarrow +\infty} \Omega =0\ ,\quad
        \lim_{\tau \rightarrow +\infty} \Sigma_+ = \alpha\ ,\quad
        \text{$0 \leq \alpha < 1$.}
\ee
\end{lemma}

\paragraph{Proof.}
Suppose that both hold. Then (\ref{Friedmann})
implies that
$
          \lim_{\tau \rightarrow +\infty} \R  = 1-\alpha^2
$.
It follows from (\ref{Spbar}) and (\ref{Ombar}) that
$
        \lim_{\tau \rightarrow +\infty} \bar{\Sigma}_+ = \alpha
$,
$ 
        \lim_{\tau \rightarrow +\infty} \bar{\Omega} = 0
$.
Then (\ref{M_lim}) and (\ref{Spbar}) imply that
$
        \lim_{\tau \rightarrow +\infty}
        \bar{\Sigma}_+' = -(1+\alpha)(1-\alpha^2) \neq 0
$,
which contradicts
$
        \lim_{\tau \rightarrow +\infty}
        \bar{\Sigma}_+ = \alpha
$.
\qed
   
\
        
We begin by proving that
\be
\label{v_0}
        \lim_{\tau \rightarrow +\infty} v =0\ ,
\ee
using the evolution equations
for $\Sigma_+$ and $v$ in the form (\ref{Sp}) and (\ref{veq}), which we
repeat for convenience:
\begin{align}
        \Sigma^{\prime}_{+} &=
        -(2-q)\Sigma_{+}-(\Nm^2+\Nc^2)-\frac{4\Omega}{3+v^2}v^{2}
\label{sig}
\\
        v^{\prime} &= \frac{6}{3-v^2}\Sigma_+(1-v^2)v \ .
\label{v}
\end{align}
Equation (\ref{sig}) implies that if $\Sp(\tau_0)=0$ for some $\tau_0$,
then $\Sp'(\tau_0)<0$, and $\Sp<0$ for $\tau > \tau_0$. We conclude that
\begin{equation}
        \hbox{the sign of $\Sigma_+$ is fixed eventually.}
\label{fix}
\end{equation}
Hence $v$ is monotone eventually and since it is also bounded,
\be
        \lim_{\tau\rightarrow\infty} v = L \ ,
\label{vlim}
\ee
where $0\le L \le 1$.
Hence
\begin{equation}
        \int_{\tau_0}^{\infty} v^{\prime}\ d\tau \quad \hbox{is finite.}
\label{intv}
\end{equation}
It follows from (\ref{sig})--(\ref{v}) that $v''$ is bounded, which
implies that $v'$ is Lipschitz continuous.
We thus conclude from Lemma~\ref{lem_Lip} that
\be
        \lim_{\tau\rightarrow\infty} v' = 0\ .
\label{vp}
\ee
We want to show that $L=0$.   
To do this we assume the contrary, namely that $0 < L \leq 1$.   

\paragraph{Case 1:} $L=1$ \\
Since $v \le 1$ and is monotone eventually then $v$ is eventually
increasing.
Hence $\Sigma_+ > 0$ eventually by (\ref{v}) and (\ref{fix}).
It follows from (\ref{sig}) that $\Sigma_+^{\prime} < 0$, hence $\Sigma_+$
is monotone decreasing eventually, and so
\be
        \lim_{\tau\rightarrow\infty}\Sigma_+=\alpha
        \ ,\quad
        0\le \alpha < 1\ .
\label{alp}
\ee
It follows that
\be
        \int_{\tau_0}^{\infty}\Sp' \ d\tau \quad \hbox{is finite.}
\ee
Since $\Sp > 0$ and $L \ne 0$ it follows from (\ref{sig}) that
\be
        \Omega \leq (-\Sp')\left(\frac{3+v^2}{4v^2}\right) \leq C (-\Sp')
        \quad \text{for $\tau \geq \tau_0$,}
\ee
where $C$ is a positive constant, and $\tau_0$ is sufficiently large.
We conclude that $\int_{\tau_0}^{\infty}\Omega\ d\tau$ is
finite.
Since $\Omega'$ is bounded by (\ref{Oeq}),
$\Omega$ is Lipschitz continuous, and it follows from Lemma~\ref{lem_Lip}
that
\be
        \lim_{\tau\rightarrow\infty}\Omega=0\ .
\ee
By Lemma~\ref{lem1} we have a contradiction.

\paragraph{Case 2:} $0< L < 1$ \\
It follows from (\ref{v}), (\ref{vlim}) and (\ref{vp}) that
$
        \lim_{\tau\rightarrow\infty}\Sigma_+=0
$,
and by integrating (\ref{v}) it follows that
$
        \int_{\tau_0}^{\infty}\Sigma_+ \ d\tau
$ is finite.
Since $L > 0$, it follows from (\ref{sig}) that
\be
        \Omega \leq - [\Sp' + (2-q)\Sp ]
                \left(\frac{3+v^2}{4v^2}\right)
        \leq - C [\Sp' + (2-q)\Sp ]
        \quad \text{for $\tau \geq \tau_0$,}
\ee
where $C$ is a positive constant, and $\tau_0$ is sufficiently large.
Since $2-q$ is bounded,
we conclude that $\int_{\tau_0}^{\infty}\Omega \ d\tau$ is finite.
The rest of the proof is the same as case 1.

\

To summarize, we have established that (\ref{v_0}) holds.
The next step is to show that $\lim_{\tau \rightarrow +\infty} \Omega$   
exists.
Extending the method used in Appendix A of NHW
we integrate (\ref{bar_Omega}) to obtain
\be
        \frac{1}{2}\ln\frac{\bar{\Omega}}{\bar{\Omega}_0}
        =\int^{\tau}_{\tau_0} \bar{\Sigma}_{+}^{2}\ d\tau
                +\int^{\tau}_{\tau_0} \frac{4\Sp v^2}{3+v^2} \ d\tau
                +\int^{\tau}_{\tau_0}M \R B_{\bar{\Omega}}\ d\tau \ ,
\label{int}
\ee
where $\bar{\Omega}_0=\bar{\Omega}(\tau_0)$ and the initial time is
chosen such that $M\le 1$ for all $\tau\ge \tau_0$.
 
Following Section 3 of NHW, we introduce an auxiliary variable $Z$
according to
\be
        Z = \frac{M\R}{\sqrt{\Omega}}\ ,
\label{Zdef}
\ee
whose evolution equation is
\be
\label{Zeq}
        Z' = \left[ -1 -2(1+\Sp)\frac{\rh}{\R} \cos 2\psi
                -3\rh M \sin 2\psi - \frac{4v^2}{3+v^2} \Sp
                \right] Z \ ,
\ee
as follows from (\ref{Req}), (\ref{Meq}) and (\ref{Oeq}).
We modify $Z$ according to
\be
\label{barZ_def}
        \bar{Z} = \frac{Z}{1
                -\frac{1}{2}M(1+\Sigma_{+})\frac{\rh}{\R} \sin 2\psi}\ ,
\ee
and obtain
\be
        \bar{Z}^{\prime} = \left[-1-\frac{4\Sp v^2}{3+v^2}
        +M B_{\bar{Z}}\right]\bar{Z}\ .
\label{zp}
\ee
Since $v$ and $M$ tend to zero as $\tau \rightarrow +\infty$, equation
(\ref{zp}) implies that
\be
        \bar{Z} = \bigO(e^{(-1+\delta)\tau})
\ee
where $\delta$ can be chosen arbitrarily small.
Equation (\ref{Zdef}) and (\ref{barZ_def}) then imply that
\be
        M \R=\bigO(e^{(-1+\delta)\tau}) \quad\hbox{and hence}
        \int^{\infty}_{\tau_0}M \R B_{\bar{\Omega}}\ d\tau
        \quad \text{is finite.}
\label{int2}
\ee
From (\ref{v}) we obtain
\be
        \frac{4\Sp v^2}{3+v^2} = \frac{2(3-v^2)vv'}{3(3+v^2)(1-v^2)}
        = f'(v)\ ,
\ee
where $f(v) =  \frac16 \ln\frac{(3+v^2)^3}{(1-v^2)}$,
and hence the second integral in (\ref{int}) is given by
\be
        \int^{\tau}_{\tau_0} \frac{4\Sp v^2}{3+v^2} \ d\tau
        = f(v)-f(v_0)\ ,
\ee
where $v_0 = v(\tau_0)$.
It follows from (\ref{v_0}) that
\be
        \int^{\infty}_{\tau_0} \frac{4\Sp v^2}{3+v^2} \ d\tau
        \quad \text{is finite.}
\label{int3}
\ee
We have thus shown that the second and third
integral terms in (\ref{int}) are bounded at late times.
Since $\Omega$ is bounded above, it follows from (\ref{int})
that
\be
        I(\tau) := \int^{\tau}_{\tau_0} \bar{\Sigma}_{+}^{2}
                \ d\tau
        \quad
        \text{is bounded above for all $\tau\ge\tau_0$.}
\label{I}
\ee
Since the integrand $\bar{\Sigma}_{+}^{2}$ is non-negative, the function
$I(\tau)$ is monotonically increasing, which implies that
$\lim_{\tau \rightarrow +\infty} I(\tau)$ exists and is finite.
It follows from (\ref{int}), (\ref{int2}) and (\ref{int3}) that
$\lim_{\tau \rightarrow +\infty} \bar{\Omega}$ exists, and hence from
(\ref{Ombar}) that $\lim_{\tau \rightarrow +\infty} \Omega$ exists, i.e.
there is a constant $L$, satisfying $0 \leq L \leq 1$ such that
\be
        \lim_{\tau \rightarrow +\infty} \Omega = L\ .   
\label{Omega_L}
\ee
        
The next step is to show that $\Sp$ and $\R$ tend to zero and that $L=1$.
Since $\bar{\Sigma}_+$ and $\barR$ are bounded, it follows from
(\ref{bar_Sp}) that
$(\bar{\Sigma}_+^2)' = 2 \bar{\Sigma}_+ \bar{\Sigma}_+'$
is uniformly bounded for $\tau \geq \tau_0$. This implies that
$\bar{\Sigma}_+^2$ is Lipschitz continuous for $\tau \geq \tau_0$.
It thus follows from (\ref{I}) and Lemma~\ref{lem_Lip} that
\be
        \lim_{\tau \rightarrow +\infty} \bar{\Sigma}_+ = 0\ ,
\label{Spbar_0}
\ee
and hence, from (\ref{Spbar}) and (\ref{M_lim}), that
\be
        \lim_{\tau \rightarrow +\infty} \Sp = 0\ .
\label{Sp_0}
\ee
It now follows from (\ref{Friedmann}), (\ref{Rbar}), (\ref{M_lim}),
(\ref{Omega_L}) and (\ref{Sp_0}) that
\be
        \lim_{\tau \rightarrow +\infty} \R
        =\lim_{\tau \rightarrow +\infty} \barR
        = 1-L\ .
\label{barR_L}
\ee
The limits (\ref{M_lim}), (\ref{v_0}), (\ref{Spbar_0}), (\ref{barR_L}) and
equation (\ref{bar_Sp}) imply that
\be
        \lim_{\tau \rightarrow +\infty} \bar{\Sigma}_+' = L-1\ ,
\ee
which contradicts (\ref{Spbar_0}) unless $L=1$. Using
(\ref{barR_L}) we obtain
\be
\label{R_0}
        \lim_{\tau \rightarrow +\infty} \R = 0 \ .
\ee
It follows from (\ref{constraint}) and (\ref{v_0}) that
\be
        \lim_{\tau \rightarrow +\infty} \rh = 0 \ .
\ee
        
Next, we will show that $M^2/\R$ tends to zero.
We note that (\ref{bar_R})--(\ref{bar_M}) give
\be
        \left(\frac{\bar{M}^2}{\barR}\right)'
        = -2 \left[ 1 + 3\bar{\Sigma}_+ + 2\bar{\Sigma}_+^2
                + M(B_{\bar{M}} + B_{\barR}) \right]
                \frac{\bar{M}^2}{\barR}\ .
\ee
It then follows from (\ref{Spbar_0}) and (\ref{M_lim}) that
\be
        \lim_{\tau \rightarrow + \infty} \frac{\bar{M}^2}{\barR} =0\ ,
\ee
and hence that  
\be
\label{RM_inf}
        \lim_{\tau \rightarrow + \infty} \frac{M^2}{\R} =0\ ,\quad
        \text{or equivalently}
        \quad
        \lim_{\tau \rightarrow + \infty} \frac{\R}{M^2} =+\infty\ .
\ee

It remains to show that 
\be
        \lim_{\tau \rightarrow + \infty} \frac{\rh}{M^2} =+\infty\ .
\ee
Equation (\ref{veq}) can be written in the form
\be
\label{N_1}
        V' = 2 \Sigma_+ V\ ,
\ee
where   
\be
        V = \frac{v}{(1-v^2)^\frac13}\ .
\ee
Let
\be
\label{N_2}
        Y = \frac{V}{\barR}\ .
\ee
It follows from (\ref{Spbar}), (\ref{bar_R}) and (\ref{N_1}) that
\be
        Y' = -2(\bar{\Sigma}_+^2 + MB)Y,
\ee
where $B$ is bounded as $\tau \rightarrow +\infty$.
This equation can be integrated to yield
\be
\label{N_3}
        Y = Y_0 e^{-\int_{\tau_0}^{\tau} \bar{\Sigma}_+^2\, d\tau}
                e^{-\int_{\tau_0}^{\tau} MB\, d\tau}\ .
\ee
We know that
\be
        \lim_{\tau \rightarrow + \infty}
        \int_{\tau_0}^{\tau} \bar{\Sigma}_+^2\, d\tau
        \quad\text{is finite.}
\ee
In addition equations (\ref{bar_M}), (\ref{Mbar}) and (\ref{Spbar_0})
imply that
\be
\label{M151}
        M = \bigO(e^{(-1+\delta)\tau}), \quad
        \text{as $\tau \rightarrow +\infty$,}
\ee
where $\delta$ is an arbitrarily small positive number.
It thus follows from (\ref{N_2}) and (\ref{N_3})--(\ref{M151}) that
\be
         \lim_{\tau \rightarrow + \infty} \frac{V}{\barR} = C\ ,
\ee
which, with
equations (\ref{N_1}), (\ref{Rbar}) and (\ref{M_lim}), implies that
\be
\label{N_4}
        \lim_{\tau \rightarrow + \infty} \frac{v}{\R} = C\ .
\ee
We now write (\ref{constraint}) in the form
\be
        1- \left(\frac{\rh}{\R}\right)^2
        = \left(\frac{4\Omega}{3+v^2}\right)^2
                \left(\frac{v}{\R}\right)^2
        \ .
\ee
It follows from (\ref{Sp_0}), (\ref{R_0}) and (\ref{Friedmann}) that
$\lim_{\tau \rightarrow + \infty} \Omega=1$.
Equation (\ref{N_4}) thus implies
\be
        \lim_{\tau \rightarrow + \infty} \left(\frac{\rh}{\R}\right)^2
        = 1- \tfrac{16}{9} C^2\ ,
\ee
which is non-zero in general. Finally, since
\be
        \frac{\rh}{M^2} = \left(\frac{\rh}{\R}\right)
                        \left(\frac{\R}{M^2}\right)
\ee
it follows from (\ref{RM_inf}) that
\be
        \lim_{\tau \rightarrow + \infty} \frac{\rh}{M^2} =+\infty\ .
\ee
\qed

\section{Derivation of decay rates}\label{app:N}

In this Appendix we use centre manifold theory
(see for example, Carr 1981, and NHW, Appendix B, for a brief summary)
to derive the asymptotic decay rates of solutions
of equations (\ref{dSphat})--(\ref{dvhat}) as $\tau \rightarrow +\infty$.
The equilibrium point $(\hat{\Sigma}_+,\hat{R},\hat{v})=(0,0,0)$ is
asymptotically stable, but non-hyperbolic, having two zero eigenvalues
(those associated with $\hat{R}$ and $\hat{v}$).
There is thus a two-dimensional centre manifold of the form
\be
        \hat{\Sigma}_+ = h(\hat{R},\hat{v}).
\ee
Using the method summarized in NHW, Appendix B, we find the following
first order approximation to the centre manifold:
\be
        \hat{\Sigma}_+ = -\hat{R}^2 -\tfrac43 \hat{v}^2 + \{ \text{higher
order terms} \}\ .
\label{centre_manifold}
\ee
To determine the asymptotic behaviour on the centre manifold, we
observe that (\ref{dvhat}) can be written as
\be
\label{Vp_V}
        V' = 2 \hSp V\ ,\quad V = \frac{\hv}{(1-\hv^2)^{1/3}}\ ,
\ee
which gives
\be
        V(\tau) = V(\tau_0) \exp \left[ \int_{\tau_0}^\tau 2 \hSp(s) \ ds
                \right]\ .
\ee
Since $\lim_{\tau \rightarrow +\infty} V(\tau)=0$, we obtain
\be
        \lim_{\tau \rightarrow +\infty} \int_{\tau_0}^\tau \hSp(s) \ ds
        = - \infty\ .
\ee
Now observe that (\ref{dRhat}) gives
\be
        \hR(\tau) = \hR(\tau_0)
        \exp \left[ \int_{\tau_0}^\tau \hSp \ ds \right]
        \exp \left[ \int_{\tau_0}^\tau \hSp^2 \ ds \right]  \ ,
\ee
where the second exponential term is greater than unity, since the
integrand
$\hSp^2$ is positive.
Hence
\be
        \frac{V}{\hR^2} = \frac{V(\tau_0)}{\hR(\tau_0)^2}
        \exp \left[ -2\int_{\tau_0}^\tau \hSp^2 \ ds \right]
\label{VR}
\ee
and
\be
        \lim_{\tau \rightarrow +\infty} \frac{V}{\hR^2} = C \ ,
\label{VRC} 
\ee
i.e. the limit exists, but could be zero.
In other words, $\hv = \bigO(\hR^2)$.
So the first order approximation of the centre manifold
(\ref{centre_manifold}) becomes
\be
        \hat{\Sigma}_+ = -\hR^2 + \{ \text{higher order terms} \}\ .
\ee
We can use this information to integrate (\ref{dRhat}), obtaining
\be
        \hR^2 = \frac{1}{2\tau} + \{ \text{higher order terms} \}\ ,
\ee
and so
\be
        \hSp^2 = \frac{1}{4\tau^2} + o\left(\frac{1}{\tau^2}\right)\ ,
\ee
which leads to
\be
        \int_{\tau_0}^\tau \hSp^2 \ ds
        = C - \frac{1}{4\tau} + o\left(\frac{1}{\tau}\right)
        = C -\tfrac12\hR^2 + o(\hR^2)\ ,
\ee  
where $C$ is a positive constant.
Hence from (\ref{Vp_V}) and (\ref{VR}) we obtain
\be
        \hv = C_v (\hR^2 + \hR^4) + o(\hR^4)\ ,
\label{vhat_R}
\ee
where $C_v$ is an arbitrary constant.
We can now reapply the centre manifold procedure to
amend (\ref{centre_manifold}) by including higher order terms that are
comparable to $\hat{v}^2$:
\be
        \hat{\Sigma}_+ = -\hat{R}^2-2\hat{R}^4 -\tfrac43 \hat{v}^2 + \{
        \text{higher order terms} \}\ .
\label{centre_manifold_2}
\ee
We then solve (\ref{dRhat}) to obtain
\be
        \hat{R} = \frac{1}{\sqrt{2\tau}} \left[ 1
        - \frac{(1+\frac43 C_v^2)(\ln \tau + C_{\R})}{4\tau}
        + o(\tau^{-1})\right]\ ,
\ee
and substitute into (\ref{vhat_R}) and (\ref{centre_manifold}) to give the
asymptotic form of $\hat{v}$ and $\hat{\Sigma}_+$. The error bound
$o(\tau^{-1})$ can be improved to $\bigO( (\frac{\ln\tau}{\tau})^2 )$,
by increasing the accuracy of the centre manifold approximation (see NHW,
page 3132).

\end{spacing}
\end{document}